\documentclass[
preprint,
superscriptaddress,
showpacs,
amsmath,amssymb,
aps,
pra,
]{revtex4-1}

\pdfoutput=1

\usepackage{graphicx}
\usepackage{dcolumn}
\usepackage{bm}
\usepackage{hyperref}
\usepackage[colorinlistoftodos]{todonotes} 
\usepackage[utf8]{inputenc}

\usepackage{color}

\begin{document}

\title{Optical spectroscopy of the $5p_{3/2} \to 6p_{1/2}$ electric dipole forbidden transition in atomic rubidium.}


\author{F. Ponciano-Ojeda}
\author{C. Mojica-Casique}
\author{S. Hern\'andez-G\'omez}
\author{O. L\'opez-Hern\'andez}
\author{J. Flores-Mijangos}
\author{F. Ram\'irez-Mart\'inez} \email{ferama@nucleares.unam.mx}
\affiliation{Instituto de Ciencias Nucleares, UNAM. Circuito Exterior, Ciudad Universitaria, 04510 M\'exico City, M\'exico.}

\author{D. Sahag\'un}
\author{R. J\'auregui}
\affiliation{Instituto de F\'\i sica, UNAM, Ciudad Universitaria, 04510 M\'exico City, M\'exico.}

\author{J. Jim\'enez-Mier} \email{jimenez@nucleares.unam.mx}
\affiliation{Instituto de Ciencias Nucleares, UNAM. Circuito Exterior, Ciudad Universitaria, 04510 M\'exico City, M\'exico.}

\date{\today} 

\begin{abstract}

We present the first evidence of excitation of the $5p_{3/2} \rightarrow 6p_{1/2}$ electric dipole-forbidden transition in atomic rubidium.
The experiments were carried out in a rubidium vapor cell using Doppler-free optical-optical double-resonance spectroscopy with counter-propagating beams.
A $5s_{1/2} \rightarrow 5p_{3/2}$ electric dipole preparation step using a diode laser locked to the $F=3 \to 4$ cyclic transition of the $D2$ line in $^{85}$Rb is used to prepare the atoms in the first excited state.
This is then followed by the $5p_{3/2}\, F_2 = 4 \rightarrow 6p_{1/2}\, F_3$ dipole-forbidden excitation ($\lambda \approx 917.5 $ nm) to establish a two-photon ladder ($\Xi$) excitation scheme.
Production of atoms in the $6p_{1/2}$ excited state is verified by detection of the $421$ nm fluorescence that results from direct decay into the $5s_{1/2}$ ground state.
The polarization dependence of the relative intensities of the lines of the decay fluorescence is also investigated.
Experimental data for different polarization configurations of the light beams used in this two-photon spectroscopy are compared with the results of calculations that consider a strong atom-field coupling in the preparation step, followed by a weak electric quadrupole excitation and the blue fluorescence decay emission.
Good agreement between experiment and this three-step model is found in the case of linear-linear polarizations.
\end{abstract}

\pacs{32.70.Cs,32.70.Fw}
\maketitle


\section{Introduction}
The study of transitions beyond the electric dipole approximation, commonly referred to as ``forbidden transitions'', in the interaction between atoms and optical radiation fields has played a significant role in the in-depth study of atomic structures.
Once limited to observation in high energy contexts, such as the studies of astrophysical phenomena and plasmas \cite{Biemont1996}, these forbidden transitions are now frequently used in other areas like metrology \cite{Rooij2011}, three-wave mixing experiments \cite{Flusberg1977a,Flusberg1977b} and parity conservation experiments \cite{Bouchiat1997}.
Thanks in part to the advent of high powered continuous-wave (CW) and pulsed laser sources, weak absorption lines, such as those given by forbidden transitions, have been observed and studied in alkali-metal vapors \cite{Guena1987,Tojo2005,Bayram2000} and cold atoms \cite{Bhattacharya2003,Pires2009,Tong2009}, as well as recently with optical-optical double resonance spectroscopy \cite{Ponciano2015,Chan2016,Mojica2016}.

In particular, $p \to p $ transitions in alkali atoms are an important part of the study of forbidden transitions.
Early experiments found that the $np \to (n+1)p $ excitation, with $n $ the ground state principal quantum number, is strong enough to be produced with moderate CW laser power \cite{Bhattacharya2003}.
Also, experiments with cold atoms conclusively proved that the $5p \to 8p $ forbidden transition in rubidium is an electric quadrupole transition with no significant contribution from the magnetic dipole term \cite{Pires2009}.
Our research group recently showed \cite{Ponciano2015,Mojica2016} that  the hyperfine structure of the $6p_{3/2} $ state could be resolved in the $5p_{3/2} \to 6p_{3/2} $ forbidden spectra of room temperature rubidium atoms.
This is possible because it is necessary first to prepare atoms in the $5p_{3/2} $ first excited state, so these are naturally collinear optical-optical double resonance experiments which allow the study of atomic transitions free of Doppler broadening \cite{Demtroder2015}.
In this experimental scheme one of the laser beams prepares atomic populations in a first excited state while the second maps the resulting population distribution onto a second excited state.
Fixing the frequency of the first of the aforementioned lasers also allows velocity-selective spectra \cite{Pinard1979} to be obtained based on the direction of propagation and the wavelength difference of the second beam relative to the fixed one.
Furthermore, we also showed that this forbidden transition is very sensitive to the polarization states of both preparation and excitation lasers, and that the choice of polarization configurations allows a direct test of electric quadrupole selection rules {\it over the atomic magnetic quantum numbers} \cite{Mojica2016}.

In this paper we extend the study of forbidden transitions in room temperature rubidium atoms to the $5p_{3/2} \to 6p_{1/2}$ electric quadrupole (E2) excitation.
This is a $p \to p $ transition to a different fine structure state ($J = 1/2 $) of the $6p $ manifold, which has a different hyperfine structure and also different transition matrix elements.
We make a detailed analysis of the dependence of the E2 spectra on the relative polarization directions of the $5s_{1/2} \to 5p_{3/2} $ linearly polarized preparation beam and the linearly polarized beam responsible of the E2 transition.
The results are then interpreted by applying the model developed for the $5p_{3/2} \to 6p_{3/2}$ excitation \cite{Mojica2016}.
This model assumes that a strong preparation step establishes the relative populations of the $5p_{3/2}\, FM_F$ magnetic sublevels and that the E2 transition intensity to the different $6p_{1/2} $ hyperfine states is determined by the geometric part of the electric quadrupole transition matrix element.

\section{Experimental setup}

\begin{figure}
	\centering
	\includegraphics[width=0.65\linewidth]{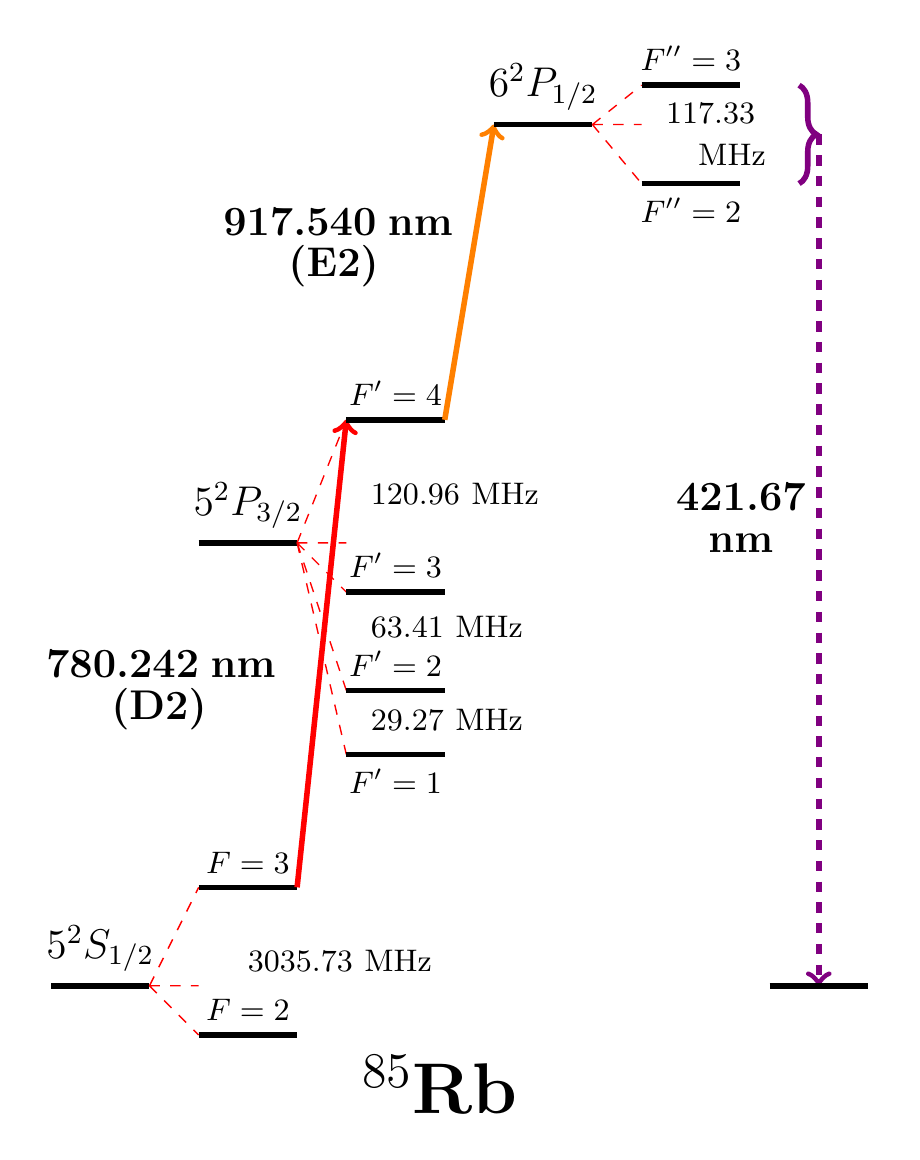}
	\caption{Energy levels of the $6p_{1/2}$ state ($^{85}$Rb). Hyperfine splittings are shown in MHz. The electric dipole excitation (D2 line) is locked to the $F \rightarrow F+1$ cyclic transition. The frequency of the electric quadrupole (E2) excitation laser is swept across the $6p_{1/2}$ hyperfine manifold in order to resolve the structure.}
	\label{fig:Rb856p}
\end{figure}

An energy level diagram showing the total angular momentum quantum numbers and hyperfine splittings for the $6p_{1/2}$ state in $^{85}$Rb is shown in figure \ref{fig:Rb856p}.
In our experiment the atoms are first prepared in the $5p_{3/2}$ excited state by an electric dipole transition, where the populations in the $M_{F'}$ magnetic sublevels reach a stationary state.
This is the same preparation step as used in \cite{Ponciano2015,Mojica2016}.
A second laser, whose frequency is swept across several hundred MHz, pumps atoms in the $5p_{3/2}$ excited state to the hyperfine $F''$ levels of the $6p_{1/2}$ state.
The hyperfine structure of the $6p_{1/2}$ state, as well as secondary cross-over features due to velocity-selective resonances, are resolved in spectra obtained by means of this excitation scheme. 
This spectroscopy, when realized with different polarization configurations of the laser beams, demonstrate the control on the preparation of the $6p_{1/2}$ hyperfine states via the E2 selection rules \cite{Mojica2016}.

The setup used for this experiment is similar to that reported in \cite{Ponciano2015}, with the main difference being the use of a commercial titanium-saphire laser \cite{MSquared} for the $917.5 $ nm electric quadrupole excitation light.
A homebuilt external cavity diode laser (ECDL)\cite{Arnold1998,Hawthorn2001} in resonance with the $5s_{1/2} \rightarrow 5p_{3/2}$ transition at $780$ nm (D2 line) is used to prepare atoms in the $5p_{3/2}$ hyperfine states.
Spontaneous decay of the excited atoms from the $6p_{1/2}$ hyperfine levels to the $5s_{1/2}$ ground state produces fluorescence at $421$ nm that is used for detection of the $5p_{3/2} \rightarrow 6p_{1/2}$ transition.
Photons produced from this decay inside the cell are collected by a system of lenses that focuses them onto the cathode of a photomultiplier tube (PMT), and the output is later processed with a lock-in amplifier to enhance the signal-noise ratio.
One expects from the theory that a doublet with the frequency splitting of the well known hyperfine structure of the $6p_{1/2}$ state \cite{Arimondo1977,Feiertag1973,Sansonetti2006} will appear in the obtained spectra.
Both lasers are linearly polarized. 
A half-wave plate in the path of the $917.5 $ nm laser is used to vary the angle between the polarization directions.

\section{Calculation of relative line intensities.}

\begin{figure}
	\centering
	\includegraphics[width=0.65\linewidth]{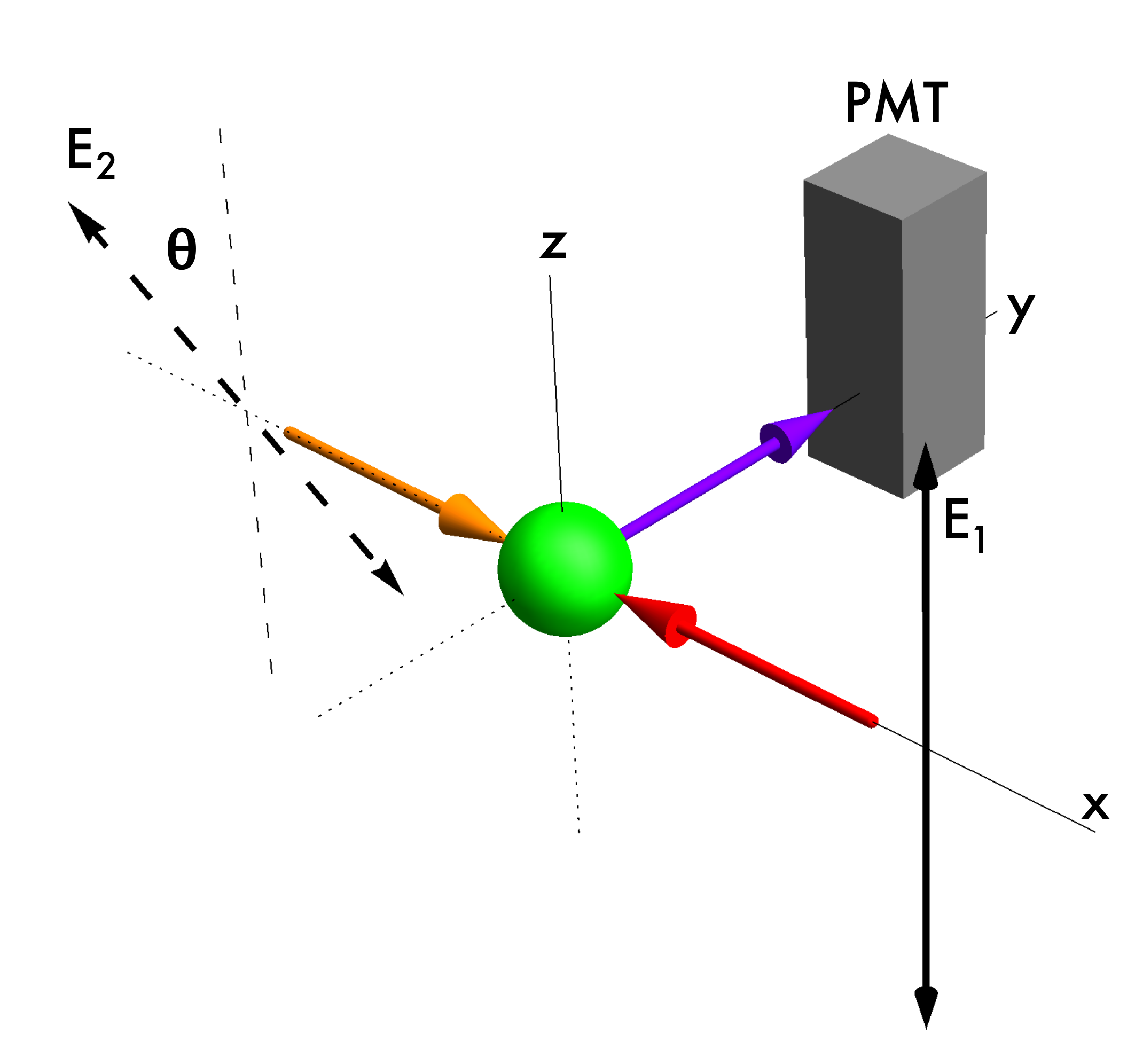}
	\caption{(Color online) Geometry used for the calculation of relative intensities of the $421$ fluorescence using the three step model presented in \cite{Ponciano2015,Mojica2016}. $\mathbf{E}_1 $ is the polarization direction of the $780 $ nm preparation laser; $\mathbf{E}_2 $ is the polarization direction of the $917.5 $ nm laser that excites the quadrupole transition. Both beams propagate along the $x $ axis and the $421 $ nm fluorescence is detected along the $y $ axis. See text for further details.}
	\label{fig:917Geometry}
\end{figure}

The overall geometry used in calculation of the relative intensities of the fluorescence resonances is shown in figure \ref{fig:917Geometry}.
The $780$ and $917.5$ beams (preparation and non-dipole excitation, respectively) propagate along the $x$ axis in opposite directions with linear polarizations along $\mathbf{E_1}$ and $\mathbf{E_2}$.
The linear polarization of the preparation laser is kept fixed along the $z$ axis. The polarization of the $917.5$ beam is rotated so that an angle $\theta$ is formed between $\mathbf{E_1}$ and $\mathbf{E_2}$. The detection of the fluorescence is performed along the positive direction of the $y$ axis by the photomultiplier tube.

The relative intensities of the observed fluorescence produced in this two-photon excitation can be calculated using the simple three-step model developed for the $5p_{3/2} \to 6p_{3/2} $ forbidden transition in references \cite{Ponciano2015} and \cite{Mojica2016}.
The probability to observe a $421$ nm photon from the decay of the $| 6p_{1/2}F_3 \rangle$ hyperfine state is given by the expression \cite{Ponciano2015,Mojica2016}
\begin{align}
	P(F_3) = \sum_{M_2,M_3,F'_1,M'_1,\lambda}& \sigma(F_2,M_2)| \langle 5p_{3/2} F_2 M_2 | T | 6p_{1/2} F_3 M_3 \rangle |^2 \nonumber \\
	&\times | \langle 6p_{1/2} F_3 M_3 | D_\lambda | 5s_{1/2} F'_1 M'_1 \rangle |^2
	\label{eqn:GeneralTransProb}
\end{align}
where $\sigma(F_2,M_2) $ is the population of the $5p_{3/2}\, F_2 M_2 $ states produced by the strong, electric dipole $5s_{1/2} \to 5p_{3/2} $ preparation step, $ \langle 5p_{3/2} F_2 M_2 | T | 6p_{1/2} F_3 M_3 \rangle |^2 $ is the weak, electric quadrupole transition probability, and $| \langle 6p_{1/2} F_3 M_3 | D_\lambda | 5s_{1/2} F'_1 M'_1 \rangle |^2 $ is the probability of decay from the $F_3 M_3 $ magnetic state of the $6p_{1/2} $ manifold.
It is important to notice that the preparation step in this experiment is the one used to study the $5p_{3/2} \to 6p_{3/2}$ transition and thus the values of the $\sigma(F_2,M_2) $  populations used in this work are the same as the ones used previously \cite{Ponciano2015,Mojica2016}.
Hyperfine and Zeeman pumping effects in this preparation step establish the magnetic state populations of the $5p_{3/2} $ hyperfine manifold which then determine the response of the atoms to the non-dipole transition.
In the transition studied here the only changes necessary are to consider the $J=1/2$ level in the $6p$ fine structure and its respective hyperfine states in the electric quadrupole excitation and the $6p_{1/2} \to 5s_{1/2} $ electric dipole decay transition matrix elements.
All the approximations made in the case of the $5p_{3/2} \rightarrow 6p_{3/2}$ transition can also be seen to hold for the present case.

It is also important to recall that different polarization configurations give place to different selections rules for the electric dipole forbidden quadrupole step.
In ref. \cite{Mojica2016} it was shown that for parallel linear polarizations we have the selection rule $\Delta M=\pm2$
while for the perpendicular case we have $\Delta M=\pm 1$.
For intermediate values of $\theta $ this model predicts spectra that are superposition of these two extremes, with weighing factors proportional to $\cos^2 \theta $ for the parallel spectrum and $\sin^2 \theta $ for the perpendicular spectrum.
Therefore the angular dependence can be written \cite{Mojica2016} in terms of the second order Legendre polynomial of $\cos \theta $. 
The fluorescence signal recorded at the ''magic angle'' ($\theta_M = \cos^{-1} \sqrt{1/3} \approx 54.7^{\circ} $) is given by $I(\theta_M) = I_\parallel + 2 I_\perp $ which is a natural normalization factor for a direct comparison between experiment and theory.

\section{Results. Comparison between experiment and theory.}

\begin{figure}
	\centering
	\includegraphics[width=0.9\linewidth]{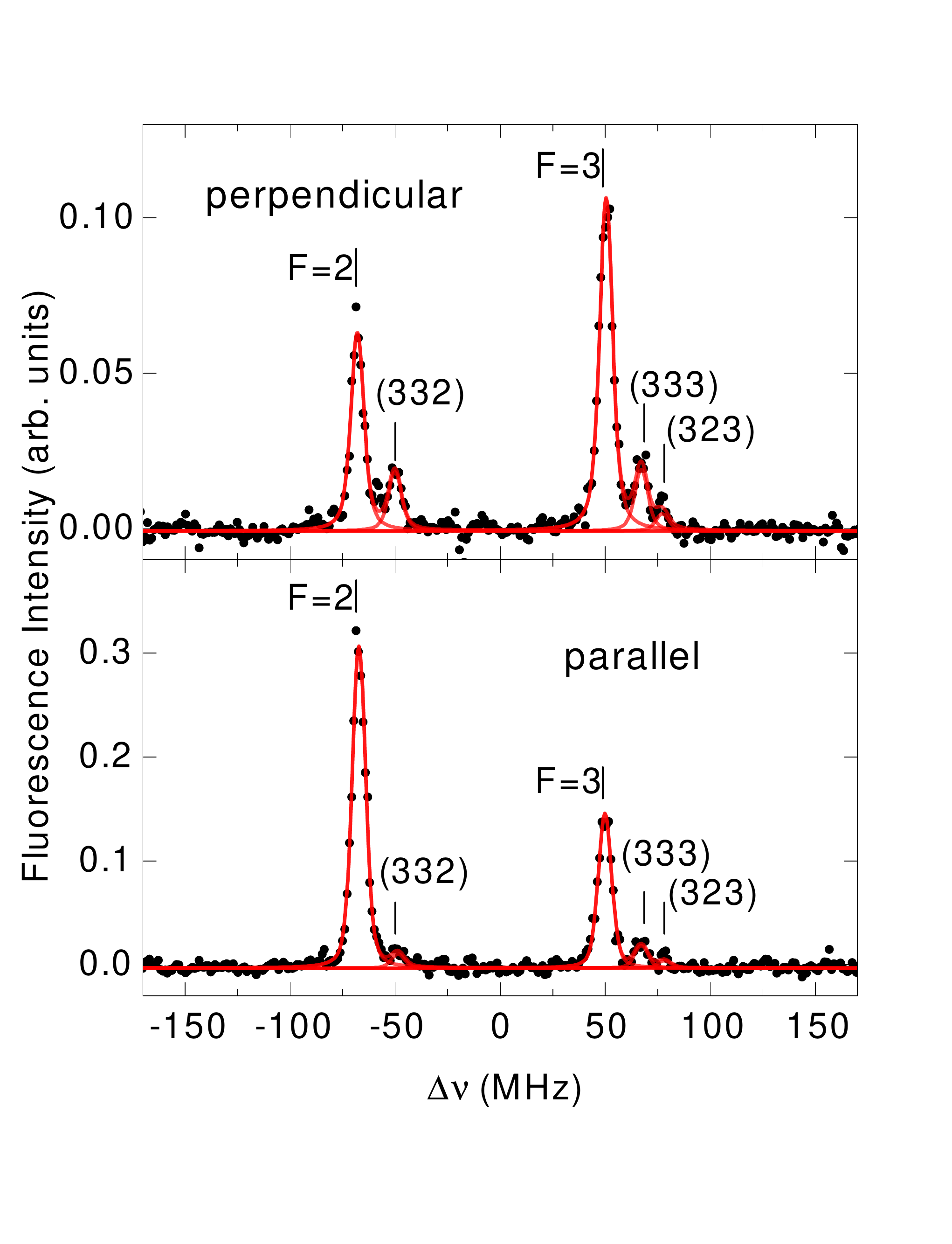}
	\caption{Spectra showing the $421$ nm decay fluorescence from the $6p_{1/2}$ state in $^{85}$Rb for parallel polarizations (bottom) and perpendicular polarizations (top). 
The center for each of the peaks, corresponding to the calculated position for each hyperfine level $F$, is indicated with vertical lines. Voigt fits for each of the main peaks are indicated by the thick (red; color online) lines, with a common width of $7.6 \pm 0.1$ MHz. The expected position of the velocity-selected peaks is also indicated.}
	\label{fig:917Spectrum}
\end{figure}

Typical spectra showing the $421 $ nm decay fluorescence as the frequency of the $917.5$ nm laser is scanned are shown in figure \ref{fig:917Spectrum}.
Here we only show spectra for parallel and perpendicular polarizations, but data were obtained for an entire angular distribution study.
The frequency scale of these plots was obtained by using the calculated hyperfine splitting for the $6p_{1/2}$ two main peaks to adjust the distance between peak centers \cite{Feiertag1973,Sansonetti2006}.
The frequency zero-reference was then shifted to the center of gravity of the $6p_{1/2}$ hyperfine manifold of the isotope in question.
A least squares fit of line profiles was done for the main observable peaks in each spectrum.
The peaks were all adjusted with the same value for the width whilst the centers and heights varied from spectrum to spectrum.
Finally, the intensity of the $F = 2 $, $3 $ spectra recorded for parallel and perpendicular polarizations were used to calculate the normalization factor 
\begin{equation}
\sum_F I(\theta_M) = \sum_F (I_\parallel + 2 I_\perp ) 
\label{norm}
\end{equation}

For the two spectra one clearly observes the two expected lines that result from an excitation sequence $5s_
{1/2}, F = 3 \rightarrow 5p_{3/2}, F = 4 \rightarrow 6p_{1/2}, F_3  $, ($F_3 = 2 $ and $3 $) for zero velocity atoms.
Due to the experiments being carried out in a room-temperature vapor cell there are also groups of atoms with non-zero velocity projections that are excited by the preparation beam.
This results in secondary peaks, which are due to the partial compensation of the Doppler shift of the preparation beam by the $917.5$ nm counter-propagating beam, and give place to the dipole-forbidden transitions appearing at a different frequency to those obtained with the maximum $F_2$ preparation \cite{Ponciano2018}.
The strongest of these velocity-selected non-dipole transitions results from the $F \rightarrow F \rightarrow F $ excitation chain ($2 \rightarrow 2 \rightarrow 2 $ in $^{87}$Rb and $3 \rightarrow 3 \rightarrow 3 $ in $^{85}$Rb).
In the $^{85}$Rb spectra small shoulders are observed at $\approx 19$ MHz above the $F = 2$, $3$ peaks, in good agreement with the position of the velocity-selected transition expected to appear at $18.1$ MHz above the zero velocity excitation.
One can also observe an additional small feature above the first velocity-selected transition for the $F = 3$ peak that corresponds to the $F \rightarrow F-1 \rightarrow F $ excitation chain, expected at $27.6$ MHz above the zero velocity excitation.
The corresponding feature for the $F = 2$ peak is not visible due to the noise.

The fits carried out on the spectra also give some information about the relative intensities of the hyperfine lines.
Using the expressions for the angular dependence of the transition probabilities given in \cite{Mojica2016} one can calculate approximate values for the relative intensities.
These values are compared to the experimental ones in table \ref{tab:CalcAngVal}.
\begin{table}
	\centering
	\begin{tabular}{|c|c|c||c|c|}
		\hline
		\multicolumn{1}{|c|}{} & \multicolumn{2}{c||}{\textbf{Parallel}} & \multicolumn{2}{c|}{\textbf{Perpendicular}}\\
		\cline{2-5}
		 & Experimental & Calculated & Experimental & Calculated\\
		\hline
		$F=2$ & 35.8 & 39.1 & 8.9 & 5.9 \\
		\hline
		$F=3$ & 19.8 & 23.1 & 13.4 & 12.9 \\
		\hline
	\end{tabular}
	\caption{Calculated and experimental percentages for the relative intensities of the hyperfine peaks in spectra for $^{85}$Rb as a function of the polarization configuration of the excitation beams. We have taken the normalization factor given by eq. \ref{norm} (see text).}
	\label{tab:CalcAngVal}
\end{table}

\begin{figure}[b]
	\centering
	\includegraphics[width=1.0\linewidth]{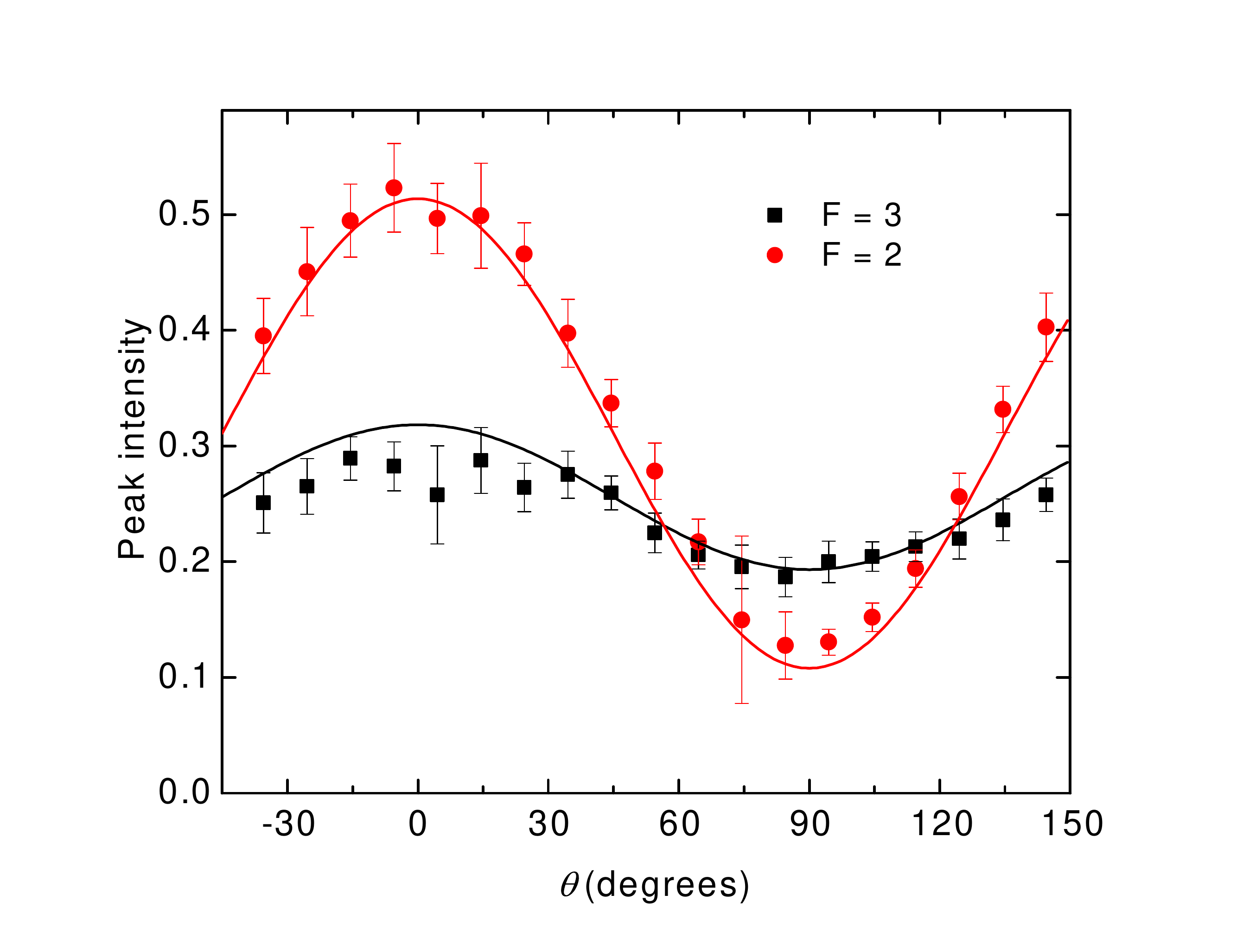}
	\caption{Angular dependence of relative intensities of the $F=3$ and $F=2$ hyperfine lines in $^{85} $Rb. The experimental data are compared with the results of the three step model calculation using the expressions obtained in \cite{Mojica2016}.
	\label{fig:AngDistr}}
\end{figure}

The result of the measurements of the angular dependence of the fluorescence line intensities as functions of the angle $\theta $ between polarization directions is shown in figure \ref{fig:AngDistr}.
The figure also shows the results of the angular distribution calculated using the three step model \cite{Mojica2016}.
For this comparison the experimental data were normalized to the total intensity of peaks $F =2 $ and $3 $ (eq. \ref{norm}).
Also, a common background was subtracted from all spectra.
As for the $5p_{3/2} \to 6p_{3/2} $ electric quadrupole transition \cite{Mojica2016}, there is very good agreement between experiment and theory.
The maxima in Fig. \ref{fig:AngDistr} for the electric quadrupole transition into both hyperfine states occur for parallel polarizations, with minima for perpendicular polarizations. 
Both experiment and theory give a much stronger angular dependence of the transition to the $F = 2 $ state.
These results confirm the validity of the theoretical model, and provide an indication of the ability to use the polarization of the light to control the population of hyperfine states excited by electric quadrupole transitions.

\section{Conclusions}
We have presented a new $p \to p $ electric dipole forbidden transition in thermal rubidium atoms, namely, the $5p_{3/2} \to 6p_{1/2} $ electric quadrupole transition.
It was also demonstrated that the three step model is a very good approximation of the excitation dynamics and plays a key role in the correct interpretation of the experimental results.
Both experimental data and the model show that the relative intensities of the forbidden spectra strongly depend on the angle $\theta$ between the linear polarization directions of preparation and excitation beams.
The model shows that this is mainly the result of different population distributions among the magnetic sublevels of both $5p_{3/2} $ and $6p_{1/2} $ excited states.
The results obtained serve as complementary information to the $5p_{3/2} \rightarrow 6p_{3/2}$ transition reported in refs. \cite{Ponciano2015,Mojica2016}, thus giving a more complete picture of the $5p_{3/2} \rightarrow 6p_J$ electric dipole-forbidden transitions in atomic rubidium.

\begin{acknowledgments}
We thank J. Rangel for his help in the construction of the diode laser. This work was supported by DGAPA-UNAM, M\'exico, under projects PAPIIT Nos. IN116309, IN110812, and IA101012, by CONACyT, M\'exico, under Basic Research project No. 44986 and National Laboratory project LN260704.
\end{acknowledgments}

\bibliography{Obs917Brief}

\end{document}